\documentclass{article}%
\usepackage{amssymb}
\usepackage{amsmath}
\usepackage{amsfonts}
\usepackage{graphicx}%

\begin{document}

\author{Vladimir K. Petrov\thanks{ E-mail address: vkpetrov@yandex.ru}}
\title{\bigskip Asymptotic transition from Fourier series to integrals in LGT}
\date{\textit{N. N. Bogolyubov Institute for Theoretical Physics}\\
\textit{\ National Academy of Science of Ukraine}\\
\textit{\ 252143 Kiev, Ukraine. }\textrm{10.10.2006}}
\maketitle

\begin{abstract}
It is shown that in asymptotic transition from Fourier series to integrals an
error and ambiguity may arise. Ambiguity reduces to a possibility of addition
of some distribution to the result. Properties of such distributions are
studied and conditions are established under which ambiguity doesn't arise.
Method for correction computation is suggested and conditions for correction
turning to zero are specified.

\end{abstract}

\section{Introduction}

In a continuum transition in a gauge theory on a lattice with length $N_{\mu}$
in a $\mu=0,1,2,3$ directions and periodic border conditions, Fourier series
in a discrete variable $\varphi_{\mu}=2\pi n_{\mu}/N_{\mu}$ ($n_{\mu}=0,...,$
$N_{\mu}-1$) must transform into series in a continual variable $0\leq
\varphi_{\mu}<2\pi$ \cite{Cr,M-M}. In a continuum limit $N_{\mu}%
\rightarrow\infty$ lattice spacings $a_{\mu}$ in $\mu\ $directions turn into
zero and physical lengths $a_{\mu}N_{\mu}=L_{\mu}$ remain finite. In a
thermodynamical limit $L_{1}L_{2}L_{3}\rightarrow\infty$, Fourier series in a
continual variable must transform into Fourier integrals. Formal change series
for integrals is applicable only for "smooth enough" functions. Although the
lattice action include only regular functions, in a limit $a_{\mu}%
\rightarrow0$ they become, in a general case, distributions. This may appear
not only in gluodynamics, but in presence mater fields as well, as it is shown
by fermion determinant computation on extremely anisotropic lattice \cite{me
cond}. In this paper we consider transitions from Fourier series to integrals
for tempered distributions $F\left(  x\right)  \in S^{\prime}$ (which satisfy
some additional restrictions specified below) that are defined as functional
$\left\langle F\left(  x\right)  ,\phi\left(  x\right)  \right\rangle $ on
fast decreasing test functions $\phi\left(  x\right)  \in S$ and that may be
represented as finite order derivatives of $F\left(  x\right)  =G^{\left(
n\right)  }\left(  x\right)  ,\ n<\infty$ of some tempered functions
$\left\vert G\left(  x\right)  \right\vert <\left\vert x\right\vert ^{\sigma
},\ x\rightarrow\pm\infty,\ \sigma<\infty$
\cite{gel-shil,vladimirov,bremermann}$.$Fourier transform of any tempered
distribution is a tempered distribution as well.

A transition from series to integrals requires an analytical continuation
function of a discrete variable to continual one, that is unique only for
functions which satisfy set of conditions of a Carlson theorem version
\cite{bieber,hille}. However, in a case of lattice gauge theories (LGT) such
conditions may appear to be too restrictive. Therefore we shall determine what
class function may introduce an ambiguity in a general case and conditions
under which no ambiguity appears will be specified.

\section{Transition from Fourier series to Fourier integral}

The transition Fourier series into integral with the infinite increasing of
period is used in \cite{Schwartz} for the definition of the Fourier integral.
One starts from locally summable function $\digamma\left(  x\right)  $ and
constructs a truncated function $\overline{F}\left(  x\right)  $ as
\begin{equation}
\overline{F}\left(  x\right)  \equiv\left\{
\begin{array}
[c]{ccc}%
\digamma\left(  x\right)  &  & -\pi\tau<x<\pi\tau\\
0 &  & otherwise
\end{array}
\right.  \label{trunc}%
\end{equation}
after that function $\overline{F}\left(  x\right)  $ is extended periodically
(with the period $2\pi\tau$) along the whole real axis $x$ by the change
$\overline{F}\left(  x\right)  \rightarrow\widetilde{F}\left(  x/\tau\right)
$ where $\widetilde{F}\left(  \varphi\right)  $ is the periodic
function\footnote{Further a tilde marks periodic functions with the period
$2\pi.$
\par
{}} with the period $2\pi$ and
\begin{equation}
\widetilde{F}\left(  x/\tau\right)  =\digamma\left(  x\right)  =\overline
{F}\left(  x\right)  ; \label{def}%
\end{equation}
for $-\pi\tau<x<\pi\tau$. Periodic function $\widetilde{F}\left(
x/\tau\right)  $ may be represented as the Fourier series
\begin{equation}
\widetilde{F}\left(  x/\tau\right)  =\sum_{n=-\infty}^{\infty}F_{n}%
\exp\left\{  inx/\tau\right\}  \label{F-s-tau}%
\end{equation}
where Fourier coefficients $F_{n}$ are given by the standard expression
\begin{equation}
F_{n}=\frac{1}{2\pi\tau}\int_{-\pi\tau}^{\pi\tau}\widetilde{F}\left(
x/\tau\right)  \exp\left\{  -inx/\tau\right\}  dx \label{p-fc}%
\end{equation}

After \cite{Schwartz} we split the region of summation $-\infty<n<\infty$ into
intervals ('bursts') $\tau k-\tau\delta k/2<n<\tau k+\tau\delta k/2$ and
\textit{approximately }change\footnote{Although term 'approximately' is not
specified in \cite{Schwartz}, one may interpret it so that the approximate
equality is assumed to become an exact one with simultaneous decreasing
$\delta k$ and $1/\tau$.} the sum in each interval for its value in the
midpoint%
\begin{equation}
\frac{1}{\tau\delta k}\sum_{n=\tau k-\tau\delta k/2}^{\tau k+\tau\delta
k/2}\exp\left\{  inx/\tau\right\}  F_{n}\simeq\left[  \exp\left\{
inx/\tau\right\}  F_{n}\right]  _{n=\tau k}=\exp\left\{  ikx\right\}
F_{k\tau} \label{smoo}%
\end{equation}

After formal change $n\rightarrow k\tau$ we get from $\left(  \ref{p-fc}%
\right)  $
\begin{equation}
\frac{1}{2\pi}\int_{-\pi\tau}^{\pi\tau}\widetilde{F}\left(  x/\tau\right)
\exp\left\{  -ikx\right\}  dx=f_{k}\tau\delta k \label{def-f2}%
\end{equation}
where it is defined
\begin{equation}
f_{k}=\tau F_{k\tau} \label{def-f}%
\end{equation}
It is assumed that after proceeding to limit $\tau\rightarrow\infty$, the
expression $\left(  \ref{def-f2}\right)  $ is converted into the inverse
Fourier transform
\begin{equation}
f_{k}=\frac{1}{2\pi}\int_{-\pi\tau}^{\pi\tau}\widetilde{F}\left(
x/\tau\right)  \exp\left\{  -ikx\right\}  dx \label{def-fk}%
\end{equation}

Although initially the small value $\delta k$ is treated as an arbitrary one,
if we claim that $\left(  \ref{smoo}\right)  $ and $\left(  \ref{def-f2}%
\right)  $ become exact equations in a limit $\tau\rightarrow\infty$, then
from $\left(  \ref{def-f2}\right)  $ we conclude, that it may happen only if
we impose the condition
\begin{subequations}
\begin{equation}
\tau\delta k=1 \label{ratio}%
\end{equation}
so that each 'burst' includes only one element between $\tau k-1/2$ and $\tau
k+1/2$ and this 'burst' contains the sole term $\exp\left\{  ikx\right\}
F_{k\tau}$. In other words, the transition from series to integral is done, in
fact, by the formal substitution
\end{subequations}
\begin{equation}
n\rightarrow k\tau;\qquad\sum_{n=-\infty}^{\infty}\rightarrow\tau\int
_{-\infty}^{\infty}dk;\qquad F_{n}\rightarrow F_{k\tau}\equiv\frac{1}{\tau
}f_{k}. \label{formal}%
\end{equation}

As it will be shown, the formal transition%
\begin{equation}
f_{k}^{formal}=\lim_{\tau\rightarrow\infty}\tau F_{k\tau} \label{exa-trans}%
\end{equation}
is admissible only under some conditions and the Fourier transform of$\ $
$f_{k}^{formal}$ may differ from $F\left(  x\right)  $, so we introduce for it
a specific notation
\begin{equation}
F^{formal}\left(  x\right)  =\int_{-\infty}^{\infty}\ f_{k}^{formal}%
\exp\left\{  ikx\right\}  dk. \label{form-x}%
\end{equation}

However, if $F\left(  x\right)  $ is computed as
\begin{equation}
F\left(  x\right)  =\lim_{\tau\rightarrow\infty}\widetilde{F}\left(
x/\tau\right)  ,
\end{equation}
the result must be true without any additional conditions on $\widetilde
{F}\left(  x/\tau\right)  $ (and consequently on $F_{k\tau}$) therefore this
result we treat as an exact one.

\section{Ambiguity introduced by the formal transition}

The formal transition is fully justified, when it used in \cite{Schwartz} for
definition of Fourier integral. When such definition is done, even for some
restricted class of functions, it may be extended even on arbitrary
distributions and it needs no reference to the original Fourier series.
One-to-one correspondence is needed only for the function and its Fourier
transform. In\textrm{\ }this paper\textrm{,} however, just the transition from
Fourier series to Fourier integral and their interrelations are of particular
interest. Hence the degree of ambiguity in such transition deserves a more
detailed consideration.

For a formal transition it is necessary to extend discrete function $F_{n}$ to
continuous one $F_{n}\rightarrow F_{t}$. We claim that $F_{n}$ may be
reconstructed uniquely from $F_{t}$, i.e. $F_{t}$ must be continuous at
integer values of argument $t=n$. Since distributions do not obligatory have a
definite value at the point, we confine ourself to such $F_{t}\in
\mathcal{S}^{\prime},$ which are locally continuous at integer values of
argument $t=n$ and denote a family of such distributions as $\mathcal{S}%
_{Z}^{\prime}$.

Recall that the distribution is locally continuous at $t=t_{0}$, if there
exists a function $\psi_{t}^{\left[  t_{0}\right]  }$ continuous in the
arbitrary small but finite vicinity of $t_{0}$ such that%
\begin{equation}
\left\langle \Psi_{t},\phi\left(  t\right)  \right\rangle =\left\langle
\psi_{t}^{\left[  t_{0}\right]  },\phi\left(  t\right)  \right\rangle
\end{equation}
for all $\phi\left(  t\right)  \in\mathcal{S}$ with a support located in this
vicinity \cite{gel-shil,vladimirov}.

Since any linear combination of distributions from $\mathcal{S}_{Z}^{\prime} $
is locally continuous at $t=n$, $\mathcal{S}_{Z}^{\prime}$ forms a linear
space as well and $\mathcal{S}_{Z}^{\prime}\subset\mathcal{S}^{\prime}$. It is
clear, that $%
\mathcal{F}%
\left[  F_{t}\right]  $ for $F_{t}\in\mathcal{S}_{Z}^{\prime}$ also forms a
linear space due to linearity of Fourier transform and we denote it as
$\mathcal{S}_{Z}^{\prime%
\mathcal{F}%
}$.

It evident, that the analytical extension $F_{n}\rightarrow F_{t}$ is not
unique in a general case, since one may add to $F_{t}$ any distribution
$\omega_{t}\in\mathcal{S}_{Z}^{\prime}$ that locally turns into zero at all
integer values of $t$. The family of such distributions $\left\{  \omega
_{t}\right\}  $ forms linear space $\Omega$.\ Indeed, $\Omega$ includes zero
element and $\alpha_{1}\omega_{1,t}+\alpha_{2}\omega_{2,t}\in\Omega$\ for any
$\omega_{k,t}\in\Omega$\ and arbitrary complex numbers $\alpha_{k}$. The
family of Fourier transforms
\begin{equation}
\omega\left(  \varphi\right)  =\int_{-\infty}^{\infty}\exp\left\{  i\varphi
t\right\}  \omega_{t}dt,\quad\omega_{t}\in\Omega
\end{equation}
forms linear space $\Omega^{%
\mathcal{F}%
}=\left\{  \omega\left(  \varphi\right)  \right\}  $.

If there are some restrictions on family of functions to which $F_{t}$ must
belong, it may lead to restrictions on $\omega_{t}$ and partially or
completely reduce ambiguity. In particular, if it is demanded that $F_{t}$
must satisfy set of conditions of a Carlson theorem version
\cite{bieber,hille} an extension $F_{n}\rightarrow F_{t}$ will be unique.
However, if one demands $F_{n}$ to be represented in a form $\left(
\ref{Fn}\right)  $ and takes this expression as an analytical extension
$F_{n}\rightarrow F_{t}$, such extension do not satisfy Carlson theorem
conditions, because in%
\begin{equation}
\overline{\lim}_{r\rightarrow\infty}\frac{1}{r}\ln\frac{\left\vert
F_{c+ir}\right\vert }{\left\vert F_{c-ir}\right\vert }\leq2\pi,\ c=const,
\end{equation}
the equality sign cannot be excluded.

Nonetheless, this extension is unique. Indeed, let us assume that one may add
to $F_{t}$ some $\omega_{t}\in\Omega$, which may be represented as%
\begin{equation}
\omega_{t}=\frac{1}{2\pi}\int_{-\pi}^{\pi}\widetilde{\omega}\left(
\varphi\right)  \exp\left\{  -i\varphi t\right\}  d\varphi. \label{om-F0}%
\end{equation}
However, if$\ \omega_{t}\ $turns into zero for all integer $t=n$ it leads to
$\widetilde{\omega}\left(  \varphi\right)  \equiv0$ and, consequently, to
$\omega_{t}\equiv0$, not only for ordinary functions, but for distributions.
In other words, there is only one distribution in $\Omega$ that may be
represented in a form $\left(  \ref{om-F0}\right)  $, and namely $\omega
_{0,t}\equiv0$.

\section{Properties of $\Omega$ and{\protect\LARGE \ }$\Omega^{%
\mathcal{F}%
}$spaces}

For a more detailed clarification of $\omega_{t}\in\Omega$ properties we need
the following procedure. Let us put in correspondence to any $F\left(
\varphi\right)  \in\mathcal{S}^{\prime}$ some distribution
\begin{equation}
\widetilde{F}\left(  \varphi\right)  \equiv\widehat{\Sigma}F\left(
\varphi\right)  \equiv\sum_{n=-\infty}^{\infty}F\left(  \varphi+2\pi n\right)
. \label{per}%
\end{equation}

Procedure $\left(  \ref{per}\right)  $, is used not infrequently (see e.g.
\cite{Schwartz}), but as a rule it is applied to functions $F\left(
\varphi\right)  $ that decrease quite rapidly with $\varphi\rightarrow
\pm\infty$. We apply it to distributions.

As it is known (see e.g. \cite{vladimirov,bremermann}) distribution $F\left(
\varphi\right)  $ may be represented as $F\left(  \varphi\right)
=F_{+}\left(  \varphi+i\varepsilon\right)  -F_{-}\left(  \varphi
-i\varepsilon\right)  $, where $\varepsilon$ is a routine positive parameter,
which is regarded as arbitrary small but finite and $F_{+}$/ $F_{-}$ are some
functions regular in the upper/lower half-plane $\varphi.$ If $F\left(
\varphi\right)  \in\mathcal{S}^{\prime}$ one can choose
\begin{equation}
F_{+}\left(  \varphi\right)  =\int_{0}^{\infty}F_{t}e^{it\varphi}dt,\qquad
F_{-}\left(  \varphi\right)  =-\int_{\infty}^{0}F_{t}e^{it\varphi}dt
\end{equation}
for Fourier integrals and%
\begin{equation}
F_{+}\left(  \varphi\right)  =\sum_{n=0}^{\infty}F_{n}e^{i\varphi n},\qquad
F_{-}\left(  \varphi\right)  =-\sum_{n=-\infty}^{-1}F_{n}e^{i\varphi n},
\end{equation}
for series, because in this case $F_{t}\in\mathcal{S}^{\prime}$ and $F_{n}$ is
discrete tempered function, so $F_{+}$/ $F_{-}$ are regular in the upper/lower
half-plane $\varphi.$

Such representation is equal to Abel-Poisson regularization of Fourier
integrals
\begin{equation}
F\left(  \varphi\right)  =%
\mathcal{F}%
\left[  F_{t},\varphi\right]  =\int_{-\infty}^{\infty}F_{t}e^{it\varphi
}dt\rightarrow\int_{-\infty}^{\infty}F_{t}e^{it\varphi-\varepsilon\left\vert
t\right\vert }dt
\end{equation}
and series
\begin{equation}
\widetilde{F}\left(  \varphi\right)  \equiv\sum_{n=-\infty}^{\infty}%
F_{n}e^{i\varphi n}\rightarrow\sum_{n=-\infty}^{\infty}F_{n}e^{i\varphi
n-\varepsilon\left\vert n\right\vert }, \label{ap}%
\end{equation}
that provides a uniform convergence which is necessary for changing the order
of integration and summation. If the opposite is not specified, it is assumed
that such regularization is done.

Writing $\left(  \ref{per}\right)  $ as
\begin{equation}
\widetilde{F}\left(  \varphi\right)  =\widehat{\Sigma}F\left(  \varphi\right)
=\sum_{n=-\infty}^{\infty}\int_{-\infty}^{\infty}e^{i\varphi t}\delta\left(
t-n\right)  F_{t}dt, \label{se}%
\end{equation}
we see that the integrand contains a product of distributions, that is
undefined in a general case. However, since we confine ourself to $F_{t}%
\in\mathcal{S}_{Z}^{\prime}$ and such $F_{t}$ are locally continuous at $t=n$
, then product $\delta\left(  t-n\right)  F_{t}$\ is always defined.
Therefore, procedure $\left(  \ref{per}\right)  $ holds to $F\left(
\varphi\right)  \in\mathcal{S}_{Z}^{\prime%
\mathcal{F}%
}$and from $\left(  \ref{se}\right)  $ we obtain%
\begin{equation}
\widetilde{F}\left(  \varphi\right)  =\sum_{n=-\infty}^{\infty}\int_{-\infty
}^{\infty}e^{i\varphi t}\delta\left(  t-n\right)  F_{t}dt=\sum_{n=-\infty
}^{\infty}e^{i\varphi n}F_{n},
\end{equation}
Hence $\widetilde{F}\left(  \varphi\right)  $ is a periodic function and
coefficients of its Fourier series are defined by a standard expression%
\begin{equation}
F_{n}=\frac{1}{2\pi}\int_{-\pi}^{\pi}\widetilde{F}\left(  \varphi\right)
\exp\left\{  -i\varphi n\right\}  d\varphi. \label{Fn}%
\end{equation}

It is easy to see that $\omega\left(  \varphi\right)  \in\Omega^{%
\mathcal{F}%
}$ if and only if
\begin{equation}
\widehat{\Sigma}\omega\left(  \varphi\right)  =\sum_{m=-\infty}^{\infty}%
\int_{-\infty}^{\infty}\exp\left\{  i\varphi t\right\}  \delta\left(
t-m\right)  \omega_{t}dt\equiv\widetilde{\omega}\left(  \varphi\right)  =0.
\label{nil}%
\end{equation}
Since $\widehat{\Sigma}$ turns into zero any $\omega\left(  \varphi\right)
\in\Omega^{%
\mathcal{F}%
}$, then $\Omega^{%
\mathcal{F}%
}$ belongs to kernel of the operator $\widehat{\Sigma}$.

Any periodic function
\begin{equation}
\widetilde{\Phi}\left(  \varphi/\tau\right)  =\sum_{n=-\infty}^{\infty}%
\phi_{n}\exp\left\{  in\varphi/\tau\right\}
\end{equation}
with the period $2\pi\tau$, where $\tau$ is not a rational number, belongs to
kernel of the operator $\widehat{\Sigma}$ as well, since
\begin{equation}
\widehat{\Sigma}\widetilde{\Phi}\left(  \varphi/\tau\right)  =\sum_{n=-\infty
}^{\infty}\phi_{n}\exp\left\{  in\varphi/\tau\right\}  \sum_{k=-\infty
}^{\infty}\delta\left(  \frac{n}{\tau}-k\right)  \equiv0
\end{equation}
If $\tau$ is equal to rational number, then $\widehat{\Sigma}\widetilde{\Phi
}\left(  \varphi/\tau\right)  \sim$ $\delta\left(  0\right)  $, hence
functions with such period do not belong to kernel of the operator
$\widehat{\Sigma}$.

Let us now introduce an operator
\begin{equation}
\widehat{\Delta}=\sum_{n=-\infty}^{\infty}\widehat{\Delta}_{n,\epsilon}%
\end{equation}
where any of operator $\widehat{\Delta}_{n,\epsilon}$ preserves only small
vicinity $\Delta_{n,\epsilon}=\left(  n-\epsilon_{n},n+\epsilon_{n}\right)
$\ of the integer $n$ in the support of distribution $\Psi_{t}\in
\mathcal{S}^{\prime}$
\begin{equation}
\quad\widehat{\Delta}_{n,\epsilon}\Psi_{t}\equiv\left\{
\begin{array}
[c]{cc}%
\Psi_{t} & t\in\Delta_{n,\epsilon}\\
0 & t\notin\Delta_{n,\epsilon}%
\end{array}
\right.  ,
\end{equation}
where $0<\epsilon_{n}\leq\epsilon$ and $\epsilon$ is an arbitrary small but
finite number.

For $F\left(  \varphi\right)  \in\mathcal{S}_{Z}^{\prime%
\mathcal{F}%
}$ we define%
\begin{equation}
\widehat{\Delta}F\left(  \varphi\right)  \equiv\int_{-\infty}^{\infty}%
\widehat{\Delta}F_{t}\exp\left\{  i\varphi t\right\}  dt=\sum_{m=-\infty
}^{\infty}\int_{m+\epsilon}^{m-\epsilon}F_{t}\exp\left\{  i\varphi t\right\}
dt. \label{FF}%
\end{equation}
Since $F_{t}$ are locally continuous at $t=m$, we can rewrite $\left(
\ref{FF}\right)  $ as
\begin{equation}
\widehat{\Delta}F\left(  \varphi\right)  =2\frac{\sin\left(  \varphi
\epsilon\right)  }{\varphi}\sum_{m=-\infty}^{\infty}F_{m}e^{i\varphi m}%
=\frac{2\sin\left(  \varphi\epsilon\right)  }{\varphi}\widetilde{F}\left(
\varphi\right)  .
\end{equation}
Therefore $\widehat{\Delta}F\left(  \varphi\right)  =O\left(  \epsilon\right)
$ for small enough $\epsilon$. At the same time%
\begin{equation}
\widehat{\Delta}\widehat{\Sigma}F\left(  \varphi\right)  =\widehat{\Delta}%
\sum_{n=-\infty}^{\infty}F\left(  \varphi+2\pi n\right)  =\widetilde{F}\left(
\varphi\right)  \sum_{n=-\infty}^{\infty}\frac{2\sin\left[  \left(
\varphi+2\pi n\right)  \epsilon\right]  }{\left(  \varphi+2\pi n\right)  }.
\label{DelF}%
\end{equation}
Applying Poisson formula for series summation, from $\left(  \ref{DelF}%
\right)  $ we obtain
\begin{equation}
\sum_{n=-\infty}^{\infty}\frac{2\sin\left(  \varphi+2\pi n\right)  \epsilon
}{\left(  \varphi+2\pi n\right)  }=\sum_{m=-\infty}^{\infty}\int_{-\infty
}^{\infty}\frac{\sin x\epsilon}{\pi x}e^{i\left(  x-\varphi\right)
m}dx=1,\quad0<\epsilon<1,
\end{equation}
which means%
\begin{equation}
\widehat{\Sigma}\widehat{\Delta}F\left(  \varphi\right)  =\widehat{\Delta
}\widehat{\Sigma}F\left(  \varphi\right)  =\widetilde{F}\left(  \varphi
\right)  .
\end{equation}

It should be noted that it is difficult to study behavior of $\widehat{\Delta
}$ for $\epsilon\rightarrow0$ in a framework of the distribution theory.
Indeed, in this limit $\left(  \widehat{\Delta},\phi\right)  \rightarrow0$ for
any test function $\phi$. In distribution theory it means $\widehat{\Delta
}\rightarrow0$. At the same time product $\widehat{\Delta}\delta\left(
\phi-n\right)  $ becomes undefined in the same limit. Therefore, $\epsilon$
will be treated always as an arbitrary small but finite parameter.

\section{General form of distributions in $\Omega$ and $\Omega^{%
\mathcal{F}%
}$}

Let us show that any $\omega\left(  \varphi\right)  \in\Omega^{%
\mathcal{F}%
}$ may be represented as
\begin{equation}
\omega\left(  \varphi\right)  =\rho\left(  \varphi+\pi\right)  -\rho\left(
\varphi-\pi\right)  \;,\quad\rho\left(  \varphi\right)  \in\mathcal{S}%
^{\prime}, \label{am}%
\end{equation}
with some restrictions on $\rho\left(  \varphi\right)  $.

Indeed, if $\omega\left(  \varphi\right)  \in\Omega^{%
\mathcal{F}%
}$, then $\rho\left(  \varphi\right)  $ may be defined as
\begin{equation}
\rho\left(  \varphi\right)  =\sum_{k=0}^{\infty}\omega\left(  \varphi
+\pi\left(  2k+1\right)  \right)  . \label{ro}%
\end{equation}
It is evident that $\left(  \ref{am}\right)  $ define $\rho\left(
\varphi\right)  $ only up to some periodic function with a period $2\pi$. On
the other hand, if $\omega\left(  \varphi\right)  $ is given by $\left(
\ref{am}\right)  $ then%
\begin{equation}
\omega_{t}=\left(  e^{i\pi t}-e^{-i\pi t}\right)  \rho_{t}, \label{om}%
\end{equation}
where%
\begin{equation}
\rho_{t}=\frac{1}{2\pi}\int_{-\infty}^{\infty}\rho\left(  \varphi\right)
\exp\left\{  -it\varphi\right\}  d\varphi.
\end{equation}
So if singularity of $\rho_{t}$ at integer $t=n$ is weak enough and do not
compensate zeros of $\left(  e^{i\pi t}-e^{-i\pi t}\right)  $ and $\left(
e^{i\pi t}-e^{-i\pi t}\right)  \rho_{t}\in\mathcal{S}_{Z}^{\prime}$, then
$\rho\left(  \varphi+\pi\right)  -\rho\left(  \varphi-\pi\right)  \in\Omega^{%
\mathcal{F}%
}$.

It follows from $\left(  \ref{nil}\right)  $ that%
\begin{equation}
\widehat{\Sigma}\omega\left(  \varphi\right)  =\widehat{\Sigma}\rho\left(
\varphi+\pi\right)  -\widehat{\Sigma}\rho\left(  \varphi-\pi\right)
=\lim_{N\rightarrow\infty}s_{N},
\end{equation}
where
\begin{align}
s_{N}  &  =\sum_{n=-N}^{N-1}\left[  \rho\left(  \varphi+\pi+2\pi n\right)
-\rho\left(  \varphi+\pi+2\pi n\right)  \right] \nonumber\\
&  =\rho\left(  \varphi+\pi-2\pi N\right)  -\rho\left(  \varphi+\pi+2\pi
N\right)  ,
\end{align}
and one may write%
\begin{equation}
\widehat{\Sigma}\omega\left(  \varphi\right)  =\lim_{N\rightarrow\infty}%
\int_{-\infty}^{\infty}\rho_{t}\left(  e^{-2it\pi N}-e^{2it\pi N}\right)
e^{i\left(  \varphi+\pi\right)  t}dt,
\end{equation}
Therefore, $\widehat{\Sigma}\omega\left(  \varphi\right)  =0$, if
\begin{equation}
\lim_{N\rightarrow\infty}\rho_{t}\exp\left\{  -it2\pi N\right\}
-\lim_{N\rightarrow\infty}\rho_{t}\exp\left\{  it2\pi N\right\}  =0.
\label{asy}%
\end{equation}

To compute $\lim_{N\rightarrow\infty}\rho_{t}\exp\left\{  \pm it2\pi
N\right\}  $ one may use an asymptotic expansion \cite{brych-prud}. In
particular, if
\begin{equation}
\rho_{t}=a_{+}\left(  t+i0\right)  ^{-\lambda_{+}}+a_{-}\left(  t-i0\right)
^{-\lambda_{-}},
\end{equation}
where $\lambda_{\pm}$ and $a_{\pm}$ are some constants, then%
\begin{equation}
\widehat{\Sigma}\omega\left(  \varphi\right)  =\left(  \frac{a_{+}%
e^{i\pi\lambda_{+}}\lim_{N\rightarrow\infty}N^{\lambda_{+}-1}}{\left(
2\pi\right)  ^{\lambda_{+}}\Gamma\left(  \lambda_{+}\right)  }-\frac
{a_{-}e^{-i\pi\lambda_{-}}\lim_{N\rightarrow\infty}N^{\lambda_{-}-1}}{\left(
2\pi\right)  ^{\lambda_{-}}\Gamma\left(  \lambda_{-}\right)  }\right)  \left(
1+O\left(  1/N\right)  \right)
\end{equation}
and we see that $\lambda_{\pm}<1$ is enough for $\widehat{\Sigma}\omega\left(
\varphi\right)  =0$.

Let us assume that some $q_{t}\in\mathcal{S}_{Z}^{\prime}$ does not turn into
zero for some integer $t=n$, so $q_{t}\notin\Omega$. It is clear that $\left(
e^{i\pi t}-e^{-i\pi t}\right)  ^{\frac{1}{m}}q_{t}\in\Omega$ for any finite
$m>0$. Since%
\begin{equation}
\lim_{m\rightarrow\infty}\left\langle \phi\left(  t\right)  ,\left(  e^{i\pi
t}-e^{-i\pi t}\right)  ^{\frac{1}{m}}q_{t}\right\rangle =\left\langle
\phi\left(  t\right)  ,q_{t}\right\rangle ,
\end{equation}
for all $\phi\in\mathcal{S}$, then space $\Omega$ is incomplete.

Let us consider as an example Fourier series with coefficients%
\begin{equation}
F_{n}=\frac{1}{n-i\sigma},\ \operatorname{Im}\sigma=0. \label{pole}%
\end{equation}
Direct extension, that is simple change $n\rightarrow t$, gives $F_{n}%
=\frac{1}{n-i\sigma}\rightarrow\frac{1}{t-i\sigma}=F_{t}$. If, however, one
takes into account%
\begin{equation}
\widetilde{F}\left(  \varphi\right)  =\sum_{n=-\infty}^{\infty}\frac
{e^{in\varphi}}{n-i\sigma}=2i\pi e^{-\sigma\varphi_{\operatorname{mod}2\pi}%
}\left(  \theta\left(  \varphi_{\operatorname{mod}2\pi}\right)  +\frac
{1}{e^{2\pi\sigma}-1}\right)  , \label{F-phi}%
\end{equation}
where $\theta\left(  t\right)  $ is the Heaviside function and use $\left(
\ref{Fn}\right)  $ for analytical extension, one gets%
\begin{equation}
F_{t}=\frac{1}{t-i\sigma}+\frac{1}{t-i\sigma}\frac{i\sin\pi t}{\sinh\pi\sigma
},
\end{equation}
so%
\begin{equation}
\omega_{t}=\frac{1}{t-i\sigma}\frac{i\sin\pi t}{\sinh\pi\sigma}.
\end{equation}
Therefore, direct extension gives for $\widetilde{F}\left(  \varphi\right)  $
expression which differs from $\left(  \ref{F-phi}\right)  $ in a function
$\omega\left(  \varphi\right)  =2\pi ie^{-\sigma\varphi}\left(  \frac
{\theta\left(  \varphi+\pi\right)  }{e^{2\sigma\pi}-1}-\frac{\theta\left(
\varphi-\pi\right)  }{1-e^{-2\sigma\pi}}\right)  $ for $-\pi\leq\varphi<\pi$.
For a finite $x$ it leads to
\begin{align}
F\left(  x\right)  -F^{formal}\left(  x\right)   &  =\lim_{\tau\rightarrow
\infty}\omega\left(  \frac{x}{\tau}\right)  =\nonumber\label{ex-cor}\\
2\pi i\lim_{\tau\rightarrow\infty}e^{-\sigma x/\tau}\left(  \frac
{\theta\left(  x/\tau+\pi\right)  }{e^{2\sigma\pi}-1}-\frac{\theta\left(
x/\tau-\pi\right)  }{1-e^{-2\sigma\pi}}\right)   &  =\frac{2\pi i}%
{e^{2\pi\sigma}-1},
\end{align}
where
\begin{equation}
F^{formal}\left(  x\right)  =\int_{-\infty}^{\infty}\frac{1}{k-i0}\exp\left\{
ikx\right\}  dk=2\pi i\theta\left(  x\right)  .
\end{equation}

In conclusion we wish to point that since $\widehat{\Delta}\widehat{\Delta
}=\widehat{\Delta}$, then $\widehat{\Delta}$ as well as $1-\widehat{\Delta} $
are projectors. So, if for some $\xi_{t}\in\mathcal{S}^{\prime}$ product
$\left(  1-\widehat{\Delta}\right)  \xi_{t}$ exists and belongs to
$\mathcal{S}^{\prime}$ at least for small enough $\epsilon$, then 
$\left(1-\widehat{\Delta}\right)  \xi_{t}\in\Omega$. Indeed, let us write
$\widehat{\Delta}$ in a form%
\begin{align}
\widehat{\Delta}  &  =\sum_{n=-\infty}^{\infty}\left(  \theta\left(
t+n+\epsilon\right)  -\theta(t+n-\epsilon)\right) \nonumber\\
&  =\sum_{m=-\infty}^{\infty}\frac{\sin\left(  2\pi m\epsilon\right)  }{\pi
m}\exp\left\{  i2\pi mt\right\}  .
\end{align}
For $R_{t}\in\mathcal{S}^{\prime}$ product $\left(  1-\widehat{\Delta
}\right)  R_{t}\in$ $\Omega$ if and only if convolution for Fourier transforms
of cofactors $%
\mathcal{F}%
\left(  1-\widehat{\Delta}\right)  \ast%
\mathcal{F}%
\left(  R_{t}|\varphi\right)  =R_{\Omega}\left(  \varphi\right)  $ belong
to$\ \Omega^{%
\mathcal{F}%
}$. In accordance with $\left(  \ref{FF}\right)  $ Fourier transform of
$\widehat{\Delta}$ may computed as
\begin{align}
\widehat{\Delta}\delta\left(  \varphi\right)   &  =\int_{-\infty}^{\infty
}\widehat{\Delta}\exp\left\{  -i\varphi t\right\}  \frac{dt}{2\pi}=%
\mathcal{F}%
\left(  \widehat{\Delta}|\varphi\right)  =\nonumber\\
&  \sum_{m=-\infty}^{\infty}\frac{\sin\left(  2\pi m\epsilon\right)  }{\pi
m}\delta\left(  \varphi-2\pi m\right)  ,
\end{align}
which leads to
\begin{equation}
R_{\Omega}\left(  \varphi\right)  =R\left(  \phi\right)  -\sum_{m=-\infty
}^{\infty}\frac{\sin\left(  2\pi m\epsilon\right)  }{\pi m}R\left(  \phi-2\pi
m\right)  .
\end{equation}
Taking into account%
\begin{equation}
\sum_{m=-\infty}^{\infty}\frac{\sin\left(  2\pi m\epsilon\right)  }{\pi
m}=1,\ 0<\epsilon<1,
\end{equation}
one can get $\widehat{\Sigma}R_{\Omega}\left(  \varphi\right)  =0$ that leads
to$\ R_{\Omega}\left(  \varphi\right)  \in\Omega^{%
\mathcal{F}%
}$. In agreement with $\left(  \ref{ro}\right)  $ we can write $R_{\Omega
}\left(  \varphi\right)  $ as
\begin{equation}
R_{\Omega}\left(  \varphi\right)  =r\left(  \varphi+\pi\right)  -r\left(
\varphi-\pi\right)  ,\ r\left(  \varphi\right)  =\sum_{k=0}^{\infty}R_{\Omega
}\left(  \varphi+\pi\left(  2k+1\right)  \right)  .
\end{equation}
It means, that $\left(  1-\widehat{\Delta}\right)  R_{t}$ $\in\ \Omega$ for
any $R_{t}\in\mathcal{S}^{\prime}$.

\section{Error introduced by the formal transition}

Applying the Poisson formula to $\left(  \ref{F-s-tau}\right)  $ we get%
\begin{align}
\widetilde{F}\left(  x/\tau\right)   &  =\sum_{m=-\infty}^{\infty}%
\int_{-\infty}^{\infty}F_{n}\exp\left\{  in\frac{x}{\tau}-2\pi imn\right\}
dn=\nonumber\\
&  \sum_{m=-\infty}^{\infty}\int_{-\infty}^{\infty}\tau F_{k\tau}\exp\left\{
ikx-2\pi imk\tau\right\}  dk.
\end{align}
Therefore, to remove an error introduced by a formal transition one must
introduce a correction%
\begin{equation}
\Xi\left(  x\right)  \equiv F\left(  x\right)  -F^{formal}\left(  x\right)
=\int_{-\infty}^{\infty}\lim_{\tau\rightarrow\infty}\sum_{m\neq0}\tau
F_{k\tau}\exp\left\{  ikx-2\pi imk\tau\right\}  dk.
\end{equation}
If $k=0$ is regular point of $\lim_{\tau\rightarrow\infty}\tau F_{k\tau}%
=f_{k}^{formal}$, then $\lim_{\tau\rightarrow\infty}\tau F_{k\tau}\exp\left\{
\pm2\pi imk\tau\right\}  =0$ for all $m\neq0,$ and we get%
\begin{equation}
\lim_{\tau\rightarrow\infty}\widetilde{F}\left(  x/\tau\right)  =\int
_{-\infty}^{\infty}f_{k}^{formal}\exp\left\{  ikx\right\}  dk=F\left(
x\right)  .
\end{equation}

Let us consider the case when an extension $F_{n}\rightarrow F_{t}$ is chosen
in such a way that $f_{k}$ appears to be singular at $k=0$. For instance, if
$F_{n}=n^{-\lambda}$ for $n>0$ and $F_{n}=0$ for $n\leq0$, one may choose
$F_{t}=t_{+}^{-\lambda}$. If regularization parameter doesn't depend on $\tau
$, it may be replaced for $+0$ and one may write $f_{k}\left(  \tau\right)
=f_{k+i0}^{\left(  +\right)  }-f_{k-i0}^{\left(  -\right)  }$, where
$f_{k}^{\left(  \pm\right)  }$ are regular in the upper/lower half-plane of a
complex variable $k$. In this case the correction may be written as
\begin{equation}
\Xi\left(  x\right)  =\lim_{\tau\rightarrow\infty}\Xi\left(  x,\tau\right)
=\lim_{\tau\rightarrow\infty}\sum_{m\neq0}F^{formal}\left(  x-2\pi
m\tau\right)  .
\end{equation}

Physical value must disappear at infinity so for them $\Xi\left(
x,\tau\right)  \rightarrow0$ with $\tau\rightarrow\infty$. For gauge fields
this condition is not mandatory and $\Xi\left(  x\right)  $ may differ from
zero at finite $x$.

With $\tau\rightarrow\infty$ correction $\Delta\left(  x\right)  $ must
disappear for any physical value $F\left(  x\right)  $. It is not true,
however, for gauge fields. For instance, if $F^{formal}\left(  x\right)
=\exp\left\{  -\left\vert x/\tau-2\pi\right\vert \right\}  $ for $\left\vert
x\right\vert <\tau$ we see that $\Xi\left(  x\right)  =\allowbreak\sinh
\frac{x}{\tau}-\coth\left(  2\pi\right)  \cosh\frac{x}{\tau}$ doesn't turn
into zero at finite $x$ and $\tau\rightarrow\infty$.

Let us finally consider the case when regularization parameter depends on
$\tau$, namely, when $f_{k}\left(  \tau\right)  $ may be represented as
\begin{equation}
f_{k}\left(  \tau\right)  =f_{k+i\sigma/\tau}^{\left(  +\right)
}-f_{k-i\sigma/\tau}^{\left(  -\right)  }, \label{Repr}%
\end{equation}
where $\sigma>0$. For $\tau\rightarrow\infty$ we get $f_{k}\left(
\tau\right)  \rightarrow f_{k}^{formal}=f_{k+i0}^{\left(  +\right)  }%
-f_{k-i0}^{\left(  -\right)  }$, hence generally $f_{k}^{formal}$ must be
treated as the distribution.

For $\Xi\left(  x\right)  =0$ it is enough that $e^{2ik\pi m\tau}f_{k\pm
i\sigma/\tau}^{\left(  \pm\right)  }\rightarrow0$ with $\tau\rightarrow\infty$
for all $m\neq0$. A general method for studying an asymptotic behavior of
$e^{ikx}f_{k\pm i0}^{\left(  \pm\right)  }$ for $x\rightarrow\pm\infty$ was
developed in \cite{br-shir}. An extension of such method to a case when
regularization parameter is finite and turns into zero only with
$\tau\rightarrow\infty$ was suggested in \cite{me A}. If $f_{k}\left(
\tau\right)  $ may be represented in a form $\left(  \ref{Repr}\right)  $,
some operation can be suggested, after which for studying an asymptotic
behavior of $e^{2ik\pi m\tau}f_{k\pm i\sigma/\tau}^{\left(  \pm\right)  }$ the
method of \cite{br-shir} can be used. Indeed, let us integrate $e^{2ik\pi
m\tau}f_{k\pm i\sigma/\tau}^{\left(  \pm\right)  }$ over real $k$. Then,
taking into account that $f_{k}^{\left(  \pm\right)  }$ are regular in the
upper/lower half-plane $k$, we may shift the integration path $\left(
-\infty,\infty\right)  \rightarrow$ $\left(  -\infty\mp i\left(  \sigma
/\tau-0\right)  ,\infty\mp i\left(  \sigma/\tau-0\right)  \right)  $ and after
changing variable $k\rightarrow k\mp i\left(  \sigma/\tau-0\right)  $ we
obtain
\begin{equation}
\frac{1}{2\pi}\int_{-\infty}^{\infty}\exp\left\{  ik2\pi m\tau\right\}
f_{k\pm i\sigma/\tau}^{\left(  \pm\right)  }dk=e^{\pm2\pi m\sigma}\frac
{1}{2\pi}\int_{-\infty}^{\infty}\exp\left\{  ik2\pi m\tau\right\}
f_{k+i0}^{\left(  \pm\right)  }dk,
\end{equation}
that allows to write symbolically
\begin{equation}
\exp\left\{  ik2\pi m\tau\right\}  f_{k\pm i\sigma/\tau}^{\left(  \pm\right)
}=\exp\left\{  ik2\pi m\tau\right\}  e^{\pm2\pi m\sigma}f_{k\pm i0}^{\left(
\pm\right)  } \label{symb}%
\end{equation}
for studying an asymptotic behavior of $\left(  \ref{symb}\right)  $ one may
apply the method developed in \cite{br-shir}. For$\ \tau\rightarrow\infty$ we
find%
\begin{equation}
e^{2i\pi m\tau k}f_{k\pm i0}^{\left(  \pm\right)  }=0,\ \pm m>0
\end{equation}
and%
\begin{equation}
e^{2i\pi m\tau k}f_{k\pm i\sigma/\tau}^{\left(  \pm\right)  }=e^{-2\pi
\left\vert m\right\vert \sigma}e^{2i\pi m\tau k}f_{k\pm i0}^{\left(
\pm\right)  }=e^{-2\pi\left\vert m\right\vert \sigma}\sum_{n=0}^{\infty}%
b_{n}^{\left(  \pm\right)  }\delta^{\left(  n\right)  }\left(  k\right)
,\ \mp m>0,
\end{equation}
where coefficients $b_{n}^{\left(  \pm\right)  }$ are defined by specific form
of $f_{k}^{\left(  \pm\right)  }$ and depend on $m\tau$. For power type
distributions $b_{n}^{\left(  \pm\right)  }$ are computed in \cite{brych-prud}%
. A general method for studying an asymptotic behavior of multidimensional
distributions was developed in \cite{vdz}. Application of this method to
studying of some specific issues of transition to continuum and thermodynamic
limit in lattice gauge series we consider in subsequent papers.

In the case considered above an expression for the correction takes the form%
\begin{equation}
\Xi\left(  x\right)  =\lim_{\tau\rightarrow\infty}\sum_{m\neq0}e^{-2\pi
\left\vert m\right\vert \sigma}F^{formal}\left(  x-2\pi m\tau\right)  .
\end{equation}

Referring again to the example $\left(  \ref{F-phi}\right)  $ we see that
$f_{k}^{formal}=\lim_{\tau\rightarrow\infty}\tau F_{k\tau}=\lim_{\tau
\rightarrow\infty}\frac{1}{k-i\sigma/\tau}=\frac{1}{k-i0}$ coincides with
direct extension $\frac{1}{n-i\sigma}\rightarrow\frac{1}{t-i\sigma}$. At the
same time, from $\left(  \ref{symb}\right)  $we get symbolic relation%
\begin{equation}
f_{k}\left(  \tau\right)  =f_{k-i\sigma/\tau}^{\left(  -\right)  }=\frac
{1}{k-i\sigma/\tau}=e^{-2\pi\left\vert m\right\vert \sigma}\frac{1}%
{k-i0}.\qquad
\end{equation}

Hence, for $\tau\rightarrow\infty$ and $m\neq0$ we get
\begin{equation}
\frac{1}{k-i0}\exp\left\{  2\pi im\tau k\right\}  =2\pi i\delta\left(
k\right)
\end{equation}
and consequently%
\begin{equation}
\lim_{\tau\rightarrow\infty}\sum_{m=-\infty}^{\infty}\exp\left\{  2\pi im\tau
k\right\}  \frac{1}{k-i\sigma/\tau}=\frac{1}{k-i0}+\frac{2\pi i}{\exp\left\{
2\pi\sigma\right\}  -1}\delta\left(  k\right)  ,
\end{equation}
that coincides with Fourier transform of $\left(  \ref{F-phi}\right)  $.

\section{Conclusions}

Studying system behavior in a finite volume with subsequent transition to
thermodynamic limit provides beneficial regularization and is used in many
domains of physics (see e.g. \cite{aoki,los}). In the present paper we show
that if in the course of regularization removing, Fourier series in LGT are
formally changed for integrals, it may introduce an ambiguity. Such ambiguity
arises due to lack of uniqueness in the analytical extension of function of
discrete argument (Fourier coefficients) to continuous one. Moreover, formal
change introduces an error, which may grow in the course of regularization
removal. All that may lead to noncoincidence in the results obtained by
different methods. Furthermore, unreasonable discrepancy may arise among gauge
theories with different expressions for actions.

It is shown that mentioned ambiguity reduced to a possibility of addition of
some distribution $\omega\left(  x/\tau\right)  \in\Omega^{%
\mathcal{F}%
}$ to the result. Properties of spaces $\Omega$ and $\Omega^{%
\mathcal{F}%
}$were studied. There are established conditions under which analytical
extension is unique even in a case when conditions of Carlson theorem are not
satisfied. A method is also suggested for the computation of correction to
remove the error, which may appear after formal transition. Conditions under
which such correction turns to zero are specified.

Author thanks G.M. Zinovjev, V.G. Kadyshevsky and A.M. Snigirev for a useful
discussions and helpful critical remarks.


\begin{thebibliography}{99}                                                                                               %


\bibitem {Cr}Creutz M, 1983 \textit{Quarks, gluons and lattices}, (Cambridge MMP)

\bibitem {M-M}Montvay I, Munster G 1997 \textit{Quantum fields on a lattice}
(Cambridge Univ. Press)

\bibitem {me cond}Petrov V K 1999 \textit{Theor.Math.Phys.}\textbf{121} 1628

\bibitem {gel-shil}Gelfand I M and Shilov G E 1964\textit{\ Generalized
functions. Properties and Operations} (Academic Press, New York).

\bibitem {vladimirov}Vladimirov V S 1979 \textit{Generalized functions in
mathematical physics} (MIR)

\bibitem {bremermann}Bremermann H 1965 \textit{Distributions, complex
variables and Fourier transforms} (Addison-Wesley)

\bibitem {bieber}Bieberbach L 1955 \textit{Analytische fortsetzung} (Springer Verlag)

\bibitem {hille}Hille E 1962 \textit{Analytic Function Theory} (Ginn and
Company, New York) volume II

\bibitem {Schwartz}Schwartz L 1966 \textit{Mathematics for the physical
sciences} (Addison-Wesley)

\bibitem {brych-prud}Brychkov Yu A  and Prudnikov A P, Integralnye preobrazovania obobschennyh funktsij, Moscow, Nauka, 1977


\bibitem {br-shir}Brychkov Yu A and Shirokov YuM 1970 \textit{Teor.Mat.Fiz.
}\textbf{4} 301, Brychkov Yu A 1970, \textit{Teor.Mat.Fiz. } \textbf{5} 98,
1973 \textit{Teor.Mat.Fiz.} \textbf{15} 375, 1975 \textit{Teor.Mat.Fiz.}
\textbf{23} 191

\bibitem {me A}Petrov V K 2004 \textit{Asymptotic series for distributions}
(Preprint hep-lat/0411031)

\bibitem {vdz}Vladimirov VS, Drozhzhinov Yu N and Zav'yalov BI 1986
\textit{Multidimensional Tauberian theorems for generalized functions }(Nauka, Moscow)

\bibitem {aoki}Aoki S et al. 2004 \textit{Phys. Rev.} \textbf{D70} 034503.

\bibitem {los}Orth B, Lippert T and Schilling K 2005 \textit{Phys.Rev}%
.\textbf{D72} 014503
\end{thebibliography}
\end{document}